\documentclass[preprint,prd,amssymb,amsmath,nobibnotes,nofootinbib]{revtex4}

\usepackage{amsfonts}
\usepackage{revsymb}
\usepackage{graphicx,epsfig}
\usepackage{placeins}

\begin{document}
\title {Scalar self-force on a static particle in Schwarzschild
using the massive field approach}
\author{Eran Rosenthal}
\affiliation{Department of Physics, University of Guelph, Guelph, Ontario,
Canada N1G 2W1}
\date{\today}

\begin{abstract}
We use the recently developed massive field approach to calculate
the scalar self-force on a static particle in a Schwarzschild
spacetime.
 In this approach the scalar self-force is obtained from the
 difference between the
 (massless) scalar field, and an auxiliary massive scalar field
 combined with a certain limiting
 process. By applying this approach to
 a static particle in Schwarzschild
  we show that the scalar self-force
 vanishes in this case. This result conforms with
 a previous analysis by Wiseman
\cite{Wiseman}.
\end{abstract}
\date{ }

\maketitle
\section {Introduction}

The self-force is a force originating from the coupling
between the charge of a
 particle and the field that this charge induces.
 In
 curved spacetime, one usually considers a fixed curved background
 spacetime and a given worldline of a charged particle. The field 
induced by the point particle diverges at the particle's location, and 
therefore a regularization method is required to calculate the correct (and finite)
self-force from this diverging field. For this problem, formal expressions for
the self-force have been derived for various types of fields. Thus,
DeWitt and Brehme derived an expression for the electromagnetic
self-force \cite{DB}. More recently Mino, Sasaki and Tanaka \cite
{MST}, and independently Quinn and Wald \cite{QW}, have derived an
expression for the gravitational self-force (here the electric charge is
replaced by the mass of the particle, and the induced electric field is 
replaced by the
linear gravitational perturbation induced by the particle's mass). Quinn
\cite{Quinn} has recently derived the corresponding expression for the scalar
self-force.

Although formal analytical expressions for the various types
 of self-forces are well known, explicit analytical calculations of
these self-forces are still a challenging task and were carried out 
only in few cases. Here, one of the main difficulties is the
non-local nature of the self-force in curved spacetime (i.e., the
self-force value at each point on the particle's trajectory
generically depends on entire past history of the particle).
For a weakly curved spacetime the above mentioned explicit self-forces expressions 
were derived by DeWitt and DeWitt
\cite{2De}, and by Pfenning an Poisson \cite{PP}.
 The self-force on a static particle was investigated analytically by several authors:
  Smith and Will have obtained a non-vanishing result for the electromagnetic
self-force on a static particle in Schwarzschild \cite{SmithWill}.
Later Lohiya \cite{Lohiya} derived the electromagnetic self-force
on a static particle for other types of background spacetimes.
Recently, Wiseman \cite{Wiseman} has showed that the scalar
self-force on a static particle in a Schwarzschild
spacetime vanishes.

A practical method to calculate the self-force for generic orbits
was devised by Barack and Ori \cite{BO} (this method was later
improved \cite{BOSCALAR,BMNOS}). In this method, one first
calculates certain regularization parameters (usually
analytically), and then uses these parameters to calculate the
self-force (usually numerically). This method was implemented
numerically in certain cases \cite{Burkoim1,Burkoim2}. In
particular, using this method (numerically) the scalar self-force
on a static particle in Kerr-Newman background was calculated by
Burko and Liu \cite {BurkoLiu}. For other approaches to the self-force problem see
\cite{Lousto,Detweiler,DW}.

Very recently, a new general method titled "the massive field
approach" for the calculation of the scalar self-force was
developed \cite{mfa}. In this paper, we implement this new method to
calculate analytically the scalar
self-force on a particle, which is held static (by some external
forces) in a Schwarzschild background spacetime (i.e., static with
respect to Schwarzschild coordinates), and thereby we show that
the self-force vanishes in this case. This result conforms with
the above mentioned analysis by Wiseman (which used a different
calculation method). The analysis given here therefore verifies
Wiseman's result and furthermore demonstrates how the
 massive field approach can be implemented for this
type of self-force calculations.

In Ref. \cite{mfa} it is shown that the scalar self-force can be obtained
from the difference between the following retarded scalar fields:
The first field is $\phi$ -- the massless scalar field induced by
the particle. This field satisfies the inhomogeneous massless
scalar field equation, with a charge density $\rho$ [see Eq.
(\ref{scalar}) below]. The second field is $\phi_\Lambda$ -- an
auxiliary massive scalar field satisfying the inhomogeneous
massive field equation, with the {\em same} charge density $\rho$
[see Eq. (\ref{scalarm}) below]. Ref. \cite{mfa} provides the
following prescription for the calculation of the scalar
self-force $f_\mu^{self}$ in terms of these fields
\begin{equation}\label{sf}
f_\mu^{self}(z_0)=q\lim_{\Lambda\rightarrow\infty}\left\{
\lim_{\delta\rightarrow 0} \Delta\phi_{,\mu}(x)+
\frac{1}{2}q\biglb[\Lambda^2n_\mu(z_0)+\Lambda
a_\mu(z_0)\bigrb]\right\}\,,
\end{equation}where,
\begin{equation}\label{deltaphi}
 \Delta\phi(x)\equiv\phi(x)-\phi_\Lambda(x)\,\,.
\end{equation}
Here, $\Lambda$ is the mass of the massive field, $z_0$ is the
self-force evaluation point on the particle's worldline, and $x$
is a point near the worldline defined as follows: at $z_0$ one
constructs a unit spatial vector $n^\mu$, which is perpendicular
to the particle's worldline but is otherwise arbitrary (i.e., at
$z_0$ the following relations are satisfied: $n^\mu n_\mu=1$,
$n^\mu u_\mu=0$). In the direction of $n^\mu$ one constructs a
geodesic, which extends out an invariant length $\delta$ to the
point $x(z_0,n^\mu,\delta)$. In the following section we shall
calculate the scalar self-force on a static particle in
Schwarzschild spacetime by implementing the mathematical
prescription given by Eq.\, (\ref{sf}).

%%%%%%%%%%%%%%%%%%% Implementation of the massive field approach %%%%%%%%%
\section{Implementation of the massive field approach}

Here we use Eq. (\ref{sf}) to calculate the scalar self-force on a
static particle in Schwarzschild. We use Schwarzschild coordinates
throughout, where $x^\alpha=(t,r,\theta,\varphi)$,
$g_{\mu\nu}=diag(-f,f^{-1},r^2,r^2\sin^2\theta)$, $f\equiv
1-\frac{2M}{r}$. We denote the particle's worldline with
$z(\tau)$, where $\tau$ denotes the particle's proper time. More explicitly
the particle worldline is characterized by the spatial coordinates
$(r_0,\theta_0,\varphi_0)$ which are constants, where $r_0>2M$.

By virtue of the spherical symmetry of the problem the two angular
components of the self-force
$f^{self}_\theta$ and $f^{self}_{\varphi}$ vanish. To calculate
the time component of the self-force $f^{self}_t$ we use Eq.
(\ref{sf}). For a static particle in Schwarzschild the scalar
fields $\phi$ and $\phi_\Lambda$ are time independent, and
therefore the term $\Delta \phi_{,t}$ in Eq. (\ref{sf}) vanishes.
Noting that the vectors $n_\mu$
 and $a_\mu$ are perpendicular to the particle's worldline
we find that $f^{self}_t$ vanishes, by virtue of Eq. (\ref{sf}).

We now consider the radial component of the self-force
$f^{self}_{r}$. Eq. (\ref{sf}) implies that the calculation of
$f^{self}_{r}$ follows from the calculation of the field
$\Delta\phi_{,r}$ near the particle's worldline, combined with a
limiting process. The first limit that has to be considered is
$\delta\rightarrow 0$. Here, we take advantage of the
arbitrariness in the definition of the unit spatial vector
$n_\mu$, and choose $n_\mu$ be the radial unit vector
$n_\mu=(0,f_0^{-1/2},0,0)$, where $f_0\equiv f(r_0)$. This choice
defines the spatial geodesic which extends out from $z_0$ to
$x$ in the direction of $n_\mu$ to be a radial spatial geodesic
(with $r\ge r_0$) denoted here with $\gamma$. The limit
$\delta\rightarrow 0$ is now equivalent to the limit $r\rightarrow
r_0$ along this geodesic. From Eq.\, (\ref{sf}) we now find that the
radial component of the self-force is given by
\begin{equation}\label{sfr}
f_r^{self}(z_0)=q\lim_{\Lambda\rightarrow\infty}\left\{
\lim_{r\rightarrow r_0} \Delta\phi_{,r}(r)+
\frac{1}{2}q\biglb[\Lambda^2n_r(z_0)+\Lambda
a_r(z_0)\bigrb]\right\}\,.
\end{equation}
Here the limit $r\rightarrow r_0$ is considered along $\gamma$.

This section is organized as follows: First, in Sec.
\ref{massless} we calculate the field $\phi_{,r}$ along the radial
geodesic $\gamma$, then in Sec. \ref{massive} we calculate an expansion
for the field $\phi_{\Lambda,r}$ in the vicinity of the particle's
worldline and along $\gamma$, and finally in Sec. \ref{selff} we
substitute $\phi_{,r}$ and $\phi_{\Lambda,r}$ in Eq.\,
(\ref{sfr}) and calculate $f_r^{self}$.

%%%%%%%%%%%%%%%%%%%%%%%%%%%%%%%%%%% Massless scalar field%%%%%%%%%%%%%%%%%%%%
\subsection{Massless scalar field}\label{massless}

Here we calculate $\phi_{,r}$ along the spatial radial geodesic
$\gamma$. The massless scalar field $\phi$ satisfies
\begin{equation}\label{scalar}
\Box\phi=-4\pi\rho\, .
\end{equation}
Here $\Box\phi\equiv{\phi_{;\mu}}^\mu$, and $\rho(x)$ is the
scalar charge density. For a point particle this 
charge density is given by 
\begin{equation}\label{density}
\rho(x)=q\int_{-\infty}^{\infty}\frac{1}{\sqrt{-g}}
\delta^4[x-z(\tau)]d\tau \,,
\end{equation}
where $g$ denotes the determinant of the background metric. For a
 static worldline Eq. (\ref{density}) gives
\begin{equation}\label{rho}
\rho=\frac{q\sqrt{f_0}\delta(r-r_0)\delta(\theta-\theta_0)
\delta(\varphi-\varphi_0)}{r^2\sin\theta}\,.
\end{equation}

Next, we solve of Eq. (\ref{scalar}) for the charge density given
by Eq. (\ref{rho}), for this purpose we decompose $\rho$ and
$\phi$ into spherical harmonics:
\begin{equation}
\rho=\sum_{l=0}^\infty\sum_{m=-l}^l\rho^{lm}(r)Y^{lm}(\theta,\varphi)\,,
\end{equation}
\begin{equation}\label{decphi}
\phi=\sum_{l=0}^\infty\sum_{m=-l}^l\phi^{lm}(r)Y^{lm}(\theta,\varphi)\,.
\end{equation}
The coefficients in the charge density decomposition are given by
\begin{equation}\label{rholm}
\rho^{lm}=q\frac{\delta(r-r_0)\sqrt{f_0}}{r^2}Y^{lm*}(\theta_0,\varphi_0)\,.
\end{equation}
The above mode decompositions together with Eq. (\ref{scalar})
give the following set of decoupled ordinary differential
equations (for a static source)
\begin{equation}
{(r^2f\phi^{lm}_{,r})}_{,r}-l(l+1)\phi^{lm}=-4\pi r^2\rho^{lm}\,.
\end{equation}
These equations have the following analytical solutions
\begin{equation}\label{philmpq}
\phi^{lm}=\frac{4\pi q
\sqrt{f_0}}{M}Y^{lm*}(\theta_0,\varphi_0)P_l(z_<)Q_l(z_>)\,.
\end{equation}
Here $z\equiv\frac{r}{M}-1$, $z_0\equiv z(r_0)$; $z_{>}$ and $z_<$
denote the larger and smaller terms from the pair $\{z_0,z\}$,
respectively; and $P_l,Q_l$ denote Legendre functions of the first
and second kind, respectively. We now substitute Eq.
(\ref{philmpq}) into Eq. (\ref{decphi}) and sum over the multipole
number $m$ which gives
\begin{equation}\label{mlesssum}
\phi=\frac{q\sqrt{f_0}}{M}
\sum_{l=0}^{\infty}(2l+1)P_l(z_<)Q_l(z_>)P_l(\cos\alpha)\,.
\end{equation}
Here,
\[
\cos\alpha\equiv
\cos\theta_0\cos\theta+\sin\theta_0\sin\theta\cos(\varphi-\varphi_0)\,.
\]
Recall that here we are interested
only in the solution along the radial geodesic $\gamma$. Along
$\gamma$ we have $\cos\alpha=1$, which considerably simplifies the summation
in Eq. (\ref{mlesssum}), and we find that along this geodesic
\[
\phi=q\frac{\sqrt{f_0}}{r-r_0}\,.
\]
Finally, we differentiate this expression with respect to $r$ and
obtain
\begin{equation}\label{phirfinal}
\phi_{,r}=- q\frac{\sqrt{f_0}}{(r-r_0)^2}\,.
\end{equation}

%%%%%%%%%%%%%%%%%%%%%%%%%%%%%%%%%Massive scalar field%%%%%%%%%%%%%%%%%%%%%%%%%%%%%%%
\subsection {Massive scalar field}\label{massive}

We now calculate an approximate expression for $\phi_{\Lambda,r}$
along the radial geodesic $\gamma$. Note that for the
implementation of the prescription summarized by Eq. (\ref{sfr}),
it is sufficient to have an approximate expression for
$\phi_{\Lambda,r}$ in the vicinity of the particle's worldline,
and as $\Lambda\rightarrow \infty$. We therefore expand
$\phi_{\Lambda,r}$ along $\gamma$ in powers of $(r-r_0)$ and in
powers of $\Lambda^{-1}$; and keep only terms that do not vanish
as $r\rightarrow r_0$, and as $\Lambda\rightarrow \infty$.

The massive scalar field $\phi_\Lambda$ satisfies
\begin{equation}\label{scalarm}
(\Box-\Lambda^2)\phi_\Lambda=-4\pi\rho\, .
\end{equation}
Here, the charge density $\rho$ is given by Eq. (\ref{rho}) -- the
same charge density as in the massless field equation. Decomposing
$\phi_\Lambda$, and $\rho$ into spherical harmonics gives
\begin{eqnarray}\label{phimdec}
&&\phi_\Lambda=\sum_{l=0}^\infty\sum_{m=-l}^l\frac{1}{r}Y^{lm}(\theta,\varphi)\phi^{lm}_\Lambda(r)\,,\\\label{rhodec}
&&\rho=\sum_{l=0}^\infty\sum_{m=-l}^l\frac{1}{r}Y^{lm}(\theta,\varphi)\tilde{\rho}^{lm}(r)\,,
\end{eqnarray}
where $\tilde{\rho}^{lm}=r{\rho}^{lm}$. From the massive field
equation (\ref{scalarm}) together with the spherical harmonics
decompositions (\ref{phimdec},\ref{rhodec}), we obtain the
following infinite set of decoupled ordinary differential
equations for the spherical harmonics coefficients
$\phi^{lm}_\Lambda$:
\begin{equation}\label{eqmassflm}
\phi^{lm}_{\Lambda,r^*r^*}-
\biglb(V^{l}(r)+\Lambda^2f\bigrb)\phi_\Lambda^{lm}=-4\pi\tilde{\rho}^{lm}f\,.
\end{equation}
Here we introduced the tortoise coordinate $r^*$ [see e.g.,
\cite{MTW}] defined by $\frac{dr*}{dr}=\frac{1}{f}$, we also defined $V^l(r)\equiv
f(\frac{f'}{r}+\frac{l(l+1)}{r^2})$ and $f'\equiv\frac{df}{dr}$.
 We now express the solution of the
inhomogeneous equation (\ref{eqmassflm})
 in terms of two linearly independent solutions of the corresponding
homogeneous equation. Denoting these homogeneous solutions with
$\chi_\pm^l$: such that $\chi_+^l$ vanishes as
$r^*\rightarrow\infty$, and $\chi_-^l$ is regular as
$r^*\rightarrow-\infty$, we find that
\begin{equation}\label{philm1}
\phi^{lm}_{\Lambda}=\frac{-4\pi q \sqrt{f_0}Y^{lm*}(\theta_0,\phi_0)}{r_0}
\frac{\chi_-^l(r_<)\chi_+^l(r_>)}{W[\chi_-^l,\chi_+^l]_{r_0}}
\end{equation}
Here
$W[\chi_-^l,\chi_+^l]={\chi_+^l}_{,r^*}{\chi_-^l}-{\chi_-^l}_{,r^*}{\chi_+^l}$
denotes the Wronskian, and $r_{>}$ and $r_<$ denote the larger and
smaller terms from the pair $\{r_0,r\}$, respectively; the
subscript $r_0$ indicates that the term in the square brackets is
evaluated at $r=r_0$.

Next, we calculate the sum over the multipole number $m$ in Eq.
(\ref{phimdec}). Introducing
$\phi_{\Lambda}^{l}\equiv\sum_{m=-l}^l\frac{1}{r}\phi_{\Lambda}^{lm}Y^{lm}\,,$
and using Eq. (\ref{philm1}) we find that along the spatial radial
geodesic $\gamma$ the multipoles $\phi_{\Lambda}^{l}$ are given by
\begin{equation}\label{phillambda}
\phi^{l}_{\Lambda}=\frac{-2qL  \sqrt{f_0}}{rr_0}
\frac{\chi_-^l(r_0)\chi_+^l(r)}{W[\chi_-^l,\chi_+^l]_{r_0}}\,,
\end{equation}
where $L\equiv l+\frac{1}{2}$.

Next, we calculate an asymptotic expansion for $\phi_{\Lambda,r}$
as $\Lambda\rightarrow \infty$. This requires summation of the
terms $\phi^{l}_{\Lambda,r}$, followed
by an asymptotic expansion in powers of $\Lambda^{-1}$.
First we consider the summation over the multipole number $l$
\begin{equation}\label{lsumphilam}
\phi_{\Lambda,r}=\sum_{l=0}^\infty\phi_{\Lambda,r}^{l}\,.
\end{equation}
Recall that for our purposes it is sufficient to calculate $\phi_{\Lambda,r}$ along the
radial geodesic $\gamma$, and up to $O[(r-r_0)^0]$ (inclusive). As
discussed in Ref.\,\cite{mfa} the field $\phi_{\Lambda,r}$
diverges on the particle's worldline. However, the individual
multipole terms $\phi_{\Lambda,r}^{l}(r_0)$, are finite on this
worldline\footnote{For an approximated expression for these
multipoles see Eq.(\ref{philamrl}) below.}, only their sum over
the multipole number $l$ diverges there. To calculate this sum
(off the worldline) it is useful to split this sum into two parts:
a sum which diverges at the limit $r\rightarrow r_0$, and a sum
which is finite at this limit. This splitting simplifies the
calculations, and allows us to use different calculation methods for the two sums.
We therefore express $\phi_{\Lambda,r}$ as
\begin{equation}\label{decphilam}
\phi_{\Lambda,r}=\sum_{l=0}^\infty
h^l+\sum_{l=0}^\infty(\phi_{\Lambda,r}^{l}-h^l)\,.
\end{equation}
This splitting is considered on the radial geodesic $\gamma$, and
the functions $h_l(r)$ will be defined below, such that the second
sum in Eq. (\ref{decphilam}) remains finite as $r\rightarrow r_0$
[see Eq. (\ref{hl}) below]. This requirement implies that the
values of $h_l(r)$ in the vicinity of the worldline must reflect
the leading asymptotic expansion of $\phi_{\Lambda,r}^{l}(r)$ as
$L\rightarrow \infty$. We comment here that a similar method for
calculating a mode-sum was devised in Ref. \cite{BOSCALAR} (for a somewhat
different purpose).

We shall now derive the required expressions for the functions $h^l(r)$. This
requires an analysis of the asymptotic behavior of
$\phi_{\Lambda,r}^{l}(r)$ as $L\rightarrow \infty$ in the
vicinity of the worldline. For this purpose we use the WKB
approximation, which enable us to calculate an asymptotic
expansion of the solutions of Eq. (\ref{eqmassflm}) in inverse
powers of $L$ (or in inverse powers of $\Lambda$). Here there is a
problem though, since the WKB approximation is invalid in the
vicinity of the black-hole horizon. This can easily be inferred
from Eq. (\ref{eqmassflm}) which has a "turning point" at the event horizon
 [i.e., the term in
the brackets in Eq. (\ref{eqmassflm}) vanishes at $r=2M$]. Therefore, we can
not impose a 
 boundary condition at $r=2M$ on the (WKB approximated) solution $\chi^l_-$.
We shall therefore consider a general approximated solution for
$\chi^l_-$  in the region where the WKB
 approximation is valid, without imposing any boundary condition.
 We shall eventually deal with this boundary condition problem in appendix A,
 by considering a different
approximation method. For the accuracy required by the
calculations below it will be sufficient to keep the first three
leading terms in the WKB approximation, which reads
\begin{eqnarray}\label{hplus}
&&\chi^{l}_+\approx e^{-S_0+S_1-S_2}\,,\\\label{hminus}
&&\chi^{l}_-\approx  c_1e^{+S_0+S_1+S_2}+c_2e^{-S_0+S_1-S_2}\,.
\end{eqnarray}
Here $c_1$ and $c_2$ are independent of $r^*$, and the functions
$S_i$, $(i=0,1,2)$ are given by [see e.g. \cite{BenderOrszag}]
\begin{eqnarray}\label{sidef}
&&S_0=\int_{2M}^rf^{-1}\sqrt{U}dr'\,,\\\nonumber
&&S_1=-\frac{1}{4}\ln U \,,\\\nonumber
&&S_2=\int_{2M}^r\frac{1}{f}\left(\frac{f\partial_{r'}(fU')}{8U^{3/2}}
-\frac{5(fU')^2}{32U^{5/2}}\right)dr'\,,
\end{eqnarray}
where $U(r')\equiv V^{l}(r')+\Lambda^2 f(r')$, and
$U'\equiv\frac{dU}{dr'}$. Examining the asymptotic properties of
the functions $\partial_r S_i$, we find that as
$L\rightarrow\infty$ (for a fixed $\Lambda$) the functions
$\partial_r S_i$ are $O(L^{1-i})$; and as
$\Lambda\rightarrow\infty$ (for a fixed $L$) the functions
$\partial_r S_i$ are $O(\Lambda^{1-i})$.
We now substitute equations (\ref{hplus},\ref{hminus}) into Eq.
(\ref{phillambda}), and differentiate with respect to $r$, this gives
\begin{equation}\label{philamrl}
\phi_{\Lambda,r}^{l}\approx
\frac{q}{\sqrt{f_0}rr_0}\frac{Le^S}{[{S_0}_{,r}+{S_2}_{,r}]_{r_0}}
(S_{,r}-\frac{1}{r})\,,
\end{equation}
where,
\begin{equation}\label{sdef}
S\equiv-[S_0(r)-S_0(r_0)]+[S_1(r)-S_1(r_0)]-[S_2(r)-S_2(r_0)]\,.
\end{equation}
This WKB approximation is accurate up to $O(L^{-1})$ (and up to
$O(\Lambda^{-1})$). At the limit where $L\rightarrow\infty$ (or
$\Lambda\rightarrow\infty$) the contributions to Eq.
(\ref{philamrl}) from the term which contains the coefficients
$c_1$ and $c_2$ [originating in Eq.\,(\ref{hminus})] vanishes
faster then any negative power of L (or $\Lambda$), and therefore
this contribution was neglected here (see Appendix A for details).

To find the asymptotic expansion for $\phi^l_{\Lambda,r}$,as
$L\rightarrow\infty$, we first calculate the leading asymptotic
expansion of $S$. Expanding $S$ gives
\begin{equation}\label{alphadef}
S=-\alpha L+O(L^0)\,.
\end{equation}
Employing equations (\ref{sidef},\ref{sdef},\ref{alphadef})
we find that $\alpha$ is approximated by
\begin{equation}\label{alphaex}
\alpha=\frac{r-r_0}{\sqrt{f_0}r_0}-\frac{(r-r_0)^2(r_0-M)}{2r_0^3f_0^{3/2}}
+O[(r-r_0)^3]\,.
\end{equation}
Eq. (\ref{alphadef}) implies that as a function of $L$ the term
$e^{S}$ in Eq. (\ref{philamrl}) has an essential singularity at
infinity. To deal with this singularity we simply multiply Eq.
(\ref{philamrl}) by $e^{-\alpha L}e^{\alpha L}$, and obtain
\begin{equation}\label{alphal}
\phi_{\Lambda,r}^{l}\approx e^{-\alpha L}\left[
\frac{q}{\sqrt{f_0}rr_0}\frac{Le^{S+\alpha
L}}{[{S_0}_{,r}+{S_2}_{,r}]_{r_0}} (S_{,r}-\frac{1}{r})\right]\,.
\end{equation}
Note that no further approximation was made here, since we merely
multiplied Eq. (\ref{philamrl}) by unity. However, in this
form the term in the square brackets does not contain an
essential singularity as $L\rightarrow\infty$, and therefore this
term can be expanded in powers of $L$ without difficulties. Note
that higher orders in the expansion of $S$ do not
give rise to a similar essential singularity, and therefore do not
require special attention. Moreover, higher orders in the
expansion of $\alpha$ in Eq. (\ref{alphaex}) will have a vanishing
contribution at the limit $r\rightarrow r_0$ (see below), and are
not required here. An expansion of the term in the square brackets
in Eq. (\ref{alphal}) reads
\begin{equation}\label{abcdef}
 \frac{q}{\sqrt{f_0}rr_0}\frac{Le^{S+\alpha
L}}{[{S_0}_{,r}+{S_2}_{,r}]_{r_0}} (S_{,r}-\frac{1}{r})=
A(r)L+B(r)+\frac{C(r)}{L}+O(L^{-2})\,.
\end{equation}
Here the coefficients $A(r)$,$B(r)$ and $C(r)$ are independent of
$L$.
We now define the functions $h^l(r)$ 
by multiplying the
 asymptotic expansion in Eq.\,(\ref{abcdef}) by $e^{-\alpha L}$, which 
gives
\begin{equation}\label{hl}
h^l(r)\equiv e^{-\alpha(r)
L}\left[A(r)L+B(r)+\frac{C(r)}{L}\right]\,.
\end{equation}
Having this definition, the functions $h^l(r)$ coincide with
the leading asymptotic expansion of $\phi_{\Lambda,r}^{l}(r)$ in
the vicinity of the worldline, as required.
Moreover, the difference
$(\phi_{\Lambda,r}^{l}-h^l)$, when evaluated on the worldline is
$O(L^{-2})$. Therefore, the second sum in Eq. (\ref{decphilam})
converges on the particle's worldline, as required. We comment
that at the limit $r\rightarrow r_0$ the coefficients
$A(r),B(r),C(r)$ coincide with the first three mode-sum
regularization parameters
 introduced by Barack and Ori in Ref. \cite{BOSCALAR}
  (for the particular problem which is considered here).

Next, we calculate the first sum in Eq. (\ref{decphilam}). Employing
Eq. (\ref{hl}) we obtain
\begin{equation}\label{hlsum}
\sum_{l=0}^\infty h^l=\sum_{l=0}^\infty e^{-\alpha
L}\left[AL+B+\frac{C}{L}\right]\,.
\end{equation}
Note that for $r>r_0$ this sum converge due to the factor
$e^{-\alpha L}$. Summing separately over the various terms
(without the coefficients) in the square brackets in Eq.
(\ref{hlsum}) gives
\begin{eqnarray}\label{sums}
&&\sum_{l=0}^\infty L e^{-\alpha
L}=\frac{\cosh(\alpha/2)}{4\sinh^2(\alpha/2)}\,,\\\nonumber
&&\sum_{l=0}^\infty  e^{-\alpha
L}=\frac{2}{\sinh(\alpha/2)}\,,\\\nonumber &&\sum_{l=0}^\infty
L^{-1}e^{-\alpha L}={2}{\textrm{arctanh}(e^{-\alpha/2})}\,.
\end{eqnarray}
Eq. (\ref{alphaex}) implies that as $r\rightarrow r_0$ the first
sum in Eq. (\ref{sums}) is $O[{(r-r_0)}^{-2}]$, the second sum is
$O[{(r-r_0)}^{-1}]$, and the third sum is
$O\{\log[{(r-r_0)}^{-1}]\}$. Our calculation has to be accurate up
to $O[(r-r_0)^0]$, and therefore the coefficients $A(r),B(r),C(r)$
in Eq. (\ref{hlsum}), and the function $\alpha(r)$ have to be
accurate up to $[O(r-r_0)^2]$. We now substitute Eq. (\ref{sums})
in Eq. (\ref{hlsum}), and calculate the coefficients $A$, $B$ and
$C$ to the required order. For this calculation we use Eqs.
(\ref{sidef},\ref{sdef},\ref{alphaex},\ref{abcdef}). We find that
the first sum in Eq. (\ref{decphilam}) is given by \cite{math}
\begin{equation}\label{fstsum}
\sum_{l=0}^\infty
h^l=-\frac{q\sqrt{f_0}}{(r-r_0)^2}+\frac{q}{2\sqrt{f_0}}\Lambda^2+O(r-r_0)\,
.
\end{equation}

Next, we consider the second sum in Eq. (\ref{decphilam}) in the
vicinity of the worldline:
\begin{equation}\label{convsum}
\sum_{l=0}^\infty[\phi_{\Lambda,r}^{l}(r)-h^l(r)]=\sum_{l=0}^\infty[\phi_{\Lambda,r}^{l}(r_0)-h^l(r_0)]+O(r-r_0)\,.
\end{equation}
Recall
  that we only need the asymptotic form (as
$\Lambda\rightarrow\infty$) of this expression. Introducing the
notations $y^l\equiv l/\Lambda$, $y^{\bar{l}}\equiv L/\Lambda$,
and $\phi_{\Lambda,r}(L,r)\equiv\phi_{\Lambda,r}^l(r)$,
$h(L,r)\equiv h^l(r)$. We find that as $\Lambda\rightarrow\infty$
the sum in Eq. (\ref{convsum}) can be approximated with a Riemann
integral
\begin{equation}\label{Rint}
\Lambda
\sum_{l=0}^\infty[\phi_{\Lambda,r}(y^{\bar{l}}\Lambda,r_0)-h(y^{\bar{l}}\Lambda,r_0)](y^{l+1}-y^l)\approx
\Lambda\int_{0}^\infty [\phi_{\Lambda,r}(\Lambda
y,r_0)-h^l(\Lambda y,r_0)] dy\,.
\end{equation}
We now substitute the WKB approximation given by Eq.
(\ref{philamrl}) into the integral in Eq. (\ref{Rint}). Note that
contributions to $\phi_{\Lambda,r}( y^{\bar{l}}\Lambda,r_0)$ from the
functions $S_i$ with $i\geq 3$ are convergent upon summation over
the multipole number $l$. However, these sums vanish as
$\Lambda\rightarrow\infty$, and therefore it is sufficient to keep
only the first three leading terms in the WKB approximation.
We now substitute Eq. (\ref{philamrl})
into  Eq. (\ref{Rint}) and expand the result in inverse powers of
$\Lambda$ [technically it is useful to substitute $U(r)=U_0+\Delta U(r)$ into
Eq. (\ref{philamrl}), where $U_0\equiv U(r_0)$, and formally
expand this expression with respect to $U_0$, and keep only terms
that will eventually give a non-vanishing contribution as
$\Lambda\rightarrow\infty$] we find that in the vicinity of the particle's worldline
the second sum in Eq. (\ref{decphilam}) is given by \cite{math}
\begin{equation}\label{scndsum}
\sum_{l=0}^\infty(\phi_{\Lambda,r}^{l}-h^l)=
\frac{qM}{2r_0^2f_0}\Lambda+O(\Lambda^{-1})+O(r-r_0)\,.
\end{equation}
Substituting equations (\ref{fstsum},\ref{scndsum}) in Eq.
(\ref{decphilam}) gives
\begin{equation}\label{philamfin}
\phi_{\Lambda,r}=-\frac{q\sqrt{f_0}}{(r-r_0)^2}+
\frac{q}{2}(\Lambda a_r +\Lambda^2n_r)+O(r-r_0)+O(\Lambda^{-1})\,.
\end{equation}
Here $n_r\equiv f_0^{-1/2}$ is the radial component of the unit
spatial vector $n_\mu$, and $a_r\equiv\frac{M}{r_0^2f_0}$ is the
radial component of the four-acceleration of the static particle
in Schwarzschild.

%%%%%%%%%%%%%%%%%%%%%%%%%%%%%% the self-force %%%%%%%%%%%%%%%%%%%%%%%%%%%%%%%%%%%%%%%%%%%%%
\subsection {The radial self-force}\label{selff}

We now calculate the radial self-force by implementing the
prescription given by Eq. (\ref{sfr}). First we calculate the
field $\Delta\phi_{,r}(r)$ in the vicinity of the particle's
worldline. From equations (\ref{phirfinal},\ref{philamfin}) we
obtain
\begin{equation}\label{difphi}
\Delta\phi_{,r}=- \frac{q}{2}(\Lambda a_r
+\Lambda^2n_r)+O(r-r_0)+O(\Lambda^{-1})\,.
\end{equation}
Note that the field $\Delta\phi_{,r}(r)$ remains finite as
$r\rightarrow r_0$ (while the fields $\phi_{,r}$ and $\phi_{\Lambda,r}$
diverge at this limit) -- this is a general property of
$\Delta\phi_{,r}(r)$ (see Ref. \cite{mfa}).
The cancellation of the divergent terms allows us to take
the limit $r\rightarrow r_0$ of $\Delta\phi_{,r}(r)$. Following
Eq. (\ref{sfr}) we add the term $\frac{q}{2}\biglb[\Lambda
a_r(z_0)+\Lambda^2n_r(z_0)\bigrb]$ to this limit. This
added term exactly annihilates the $O(\Lambda)$ and $O(\Lambda^2)$
terms in our expression -- this is a general property as well (see Ref.
 \cite{mfa}). We now complete
this calculation by taking the limit
$\Lambda\rightarrow\infty$ in Eq. (\ref{sfr}), which gives
\begin{equation}
f_r^{self}(z_0)=q\lim_{\Lambda\rightarrow\infty}\left\{
\lim_{r\rightarrow r_0} \Delta\phi_{,r}(r)+
\frac{1}{2}q\biglb[\Lambda^2n_r(z_0)+\Lambda
a_r(z_0)\bigrb]\right\}=0\,.
\end{equation}
Combining this result with the results for the other components of
the self-force, we conclude that the scalar self-force on a static
particle in Schwarzschild vanishes.

%%%%%%%%%%%%%%%%%%%%%%%%%%%%%%%%%  acknowledgments  %%%%%%%%%%%%%%%%%%%%%%%%%%%
\subsection*{Acknowledgment}
I am grateful to Amos Ori for valuable discussions.

%%%%%%%%%%%%%%%%%%%%%%%%%%%%%%%%%%%% APPENDIX A%%%%%%%%%%%%%%%%%%%%%%%%%%%%%%%
\appendix\label{appa}
\section{Green-Liouville approximation}

Here we shall derive Eq.\,(\ref{philamrl}). In particular we shall
provide justification for neglecting the term which depends on
$c_1$ and $c_2$ in the expression for $\phi_{\Lambda,r}^{l}$. As
discussed in subsection \ref{massive}, the WKB approximation is
invalid in the vicinity of the event horizon. This situation
prevents us from patching the WKB approximation for the
homogeneous solution $\chi_-^l$ to a regular boundary condition at
the event horizon. In the region where the WKB approximation is
valid, the solution $\chi_-^l$  is  given by Eq.\,(\ref{hminus}).
By substituting equations\,(\ref{hplus},\ref{hminus}) into Eq.
(\ref{phillambda}), and differentiating with respect to $r$ we
find that
\begin{equation}\label{fullphilambar}
\phi_{\Lambda,r}^{l}\approx
\frac{q}{\sqrt{f_0}rr_0}\frac{Le^S}{[{S_0}_{,r}+{S_2}_{,r}]_{r_0}}
(S_{,r}-\frac{1}{r})(1+\frac{c_2}{c_1}e^{-2S_0-2S_2})\,.
\end{equation}
We now focus on the contribution from the last term in the last
brackets, and explain why this term can be neglected here. For
this we follow an analysis by Rowan and Stephenson
\cite{RowanStephenson}, who showed that by using the
Green-Liouville approximation method, one can obtain an
approximate solution to Eq. (\ref{eqmassflm}), which is valid in the entire
$r\geq 2M$ region. First, we express the homogeneous wave equation
for $\chi^l_\pm$ as
\begin{equation}\label{eqchi}
[r^2f\chi^l_{,r}]_{,r}-[l(l+1)+\Lambda^2r^2]\chi^l=0\,.
\end{equation}
Next we introduce $x=(r/M)f$, $N=M\Lambda$; and make the following
transformations from $x$ to $\xi$:
\[
{\xi'}^2=\beta^2\biglb(\frac{2+x}{x}\bigrb)+\frac{\alpha^2}{x(x+2)}\,.
\]
Here, ${\xi'}=(\frac{d\xi}{dx})$, $\beta^2=N^2/k^2$, $\alpha^2=[l(l+1)]/k^2$, and
$k^2=N^2+l(l+1)$. Note that $0\leq \alpha \leq 1$, $0\leq \beta
\leq 1$. The new variable $\xi$ is defined by
\[
\xi=\int_0^x |\xi'| dx\, .
\]
Next, we define $\psi^l$ such that
\begin{equation}\label{psidef}
\psi^l=\sqrt{\xi' x(x+2)}\chi^l\,.
\end{equation}
Eq. (\ref{eqchi}) now reads
\begin{equation}\label{psieq}
\psi^l_{,\xi\xi}-\biglb[k^2-(2\xi)^{-2}+g_1(\xi)\bigrb]\psi^l=0\,.
\end{equation}
Here $g_1(\xi)$ is a slowly varying function of $\xi$, which is
bounded everywhere. This function is $O(1)$ near the event horizon, and it is
$O(\xi^{-2})$ as $\xi\rightarrow\infty$. The full expression for
$g_1(\xi)$ (see \cite{RowanStephenson}) is not required here.

For large values of $k$ (which corresponds to large values of $N$
and/or large values of $l$) Eq.\,(\ref{psieq}) can be solved using
perturbation analysis. The leading order in this approximation is
obtained by completely neglecting the small contribution from the
function $g_1(\xi)$ in this equation. The remaining equation
can be solved exactly, giving two linearly independent solutions,
from which the corresponding $\chi_\pm^l$ can be calculated.
 These approximate homogeneous solutions are
given by
\[
\chi^l_+\approx \sqrt{\xi}[{\xi' x(x+2)}]^{-1/2}K_0(k\xi),\,
\chi^l_-\approx \sqrt{\xi}[{\xi' x(x+2)}]^{-1/2}I_0(k\xi) \,.
\]
Here $I_0$ and $K_0$  are the modified Bessel functions of the
first and second kind, respectively. By continuing the
perturbation analysis to higher orders we find that the
approximate solution $\chi^l_-$ at any (fixed) order has the
following asymptotic form (as $k\rightarrow\infty$)
\begin{equation}\label{genexp}
\chi^l_-\approx F_-(k,\xi)e^{k \xi}+F_+(k,\xi)e^{-k \xi}\,.
\end{equation}
For $\xi\ne0$ the functions $ F_\pm(k,\xi)$  vanish at the limit
$k\rightarrow\infty$. At the domain where both WKB and
Green-Liouville approximations are valid, we equate Eq.
(\ref{genexp}) with the corresponding WKB approximation [i.e.,
with Eq. (\ref{hminus}), extended to the required order].
Using this equation (for an arbitrary fixed order) we find
that the term $\frac{c_2}{c_1}e^{-2S_0-2S_2}$ approaches zero
faster then any negative power of $k$, as $k\rightarrow\infty$. We
therefore conclude that the last term in the last brackets in
Eq.\,(\ref{fullphilambar}) can be neglected in our calculation.
Note, that the above mentioned equation can only provide us with a bound 
 (which is sufficient for our purpose). This equation, however, can not be used 
to determine the value of the coefficient $c_2$
 (though it can be used  to determine 
$c_1$ up to a given order). The difficulty with the coefficient $c_2$
is that it multiplies a subdominant term -- a
term which vanishes exponentially. 

We comment here that the entire perturbation analysis can be done
with Green-Liouville approximation, this method has the advantage
of being valid in the entire $r\ge 2M$ region. However, we find
that the WKB approximation is simpler, especially for the higher
orders which are required by our analysis.


\begin{thebibliography}{99}
\bibitem{math}
This expression was calculated with Mathematica software.
\end{thebibliography}
\end{document}